# High-Temperature Activated AB$_2$ Nanopowders for Metal Hydride Hydrogen Compression

E.D. Koultoukis[1,2,3], E.I Gkanas[1,2], S.S. Makridis[2,4,*], C. N. Christodoulou[2,5], D. Fruchart[6] and A.K. Stubos[2]

[1]*McPhy Energy S.A., Z.A. Quartier Riétière, La Motte-Fanjas, 26190, France*
[2]*Institute of Nuclear Technology and Radiation Protection, NCSR "Demokritos", Ag. Paraskevi, Athens, 15-310, Greece*
[3]*Department of Physics, Aristotle University of Thessaloniki, Egnatia Street, Thessaloniki, GR 54124, Greece*
[4]*Department of Mechanical Engineering, University of Western Macedonia, 50100, Kozani, Greece*
[5]*Hystore Technolgies LtD, 30 Spyrou Kyprianou, Ergates Industrial Area, Nicosia, 2643, Cyprus*
[6]*Laboratoire de Cristallographie du CNRS, 25 Avenue des Martyrs, BP 166, 38042 Grenoble Cedex 9, France*

**Abstract**

A reliable process for compressing hydrogen and for removing all contaminants is that of the metal hydride thermal compression. The use of metal hydride technology in hydrogen compression applications though, requires thorough structural characterization of the alloys and investigation of their sorption properties. The samples have been synthesized by induction - levitation melting and characterized by Rietveld analysis of the X-Ray diffraction (XRD) patterns. Volumetric PCI (Pressure-Composition Isotherm) measurements have been conducted at 20, 60 and 90 $^o$C, in order to investigate the maximum pressure that can be reached from the selected alloys using water of 90$^o$C. Experimental evidence shows that the maximum hydrogen uptake is low since all the alloys are consisted of Laves phases, but it is of minor importance if they have fast kinetics, given a constant volumetric hydrogen flow. Hysteresis is almost absent while all the alloys release nearly all the absorbed hydrogen during desorption. Due to hardware restrictions, the maximum hydrogen pressure for the measurements was limited at 100 bars. Practically, the maximum pressure that can be reached from the last alloy is more than 150 bars.

*correspondence should be addressed: sofmak@ipta.demokritos.gr

1. INTRODUCTION

Energy is one of the key elements for global peace and nature rescue. An increase in the energy consumption of a country provides a positive impact on the economic as well as social development of the country [1]. Furthermore, the supply, the effective utilization and the clean fuel is very crucial for every country as a factor of social, economic and institutional sustainability [2, 3]. An investigation for alternative fuels and energy technologies became important the last years for future energy stability [4].

Since it was found that hydrogen can be derived from water through electrolysis, i.e. consuming energy for separating hydrogen from oxygen, it became evident that the reverse reaction is also attainable. Hydrogen could be used for producing energy, either taking advantage of the enthalpy produced when it is combusted in a cylinder of an internal combustion engine (ICE), or directly for producing electricity in a fuel cell, for many other purposes. However, the growing technology of using other sustainable energy forms, such as wind, solar, wave, tidal is very promising, but it cannot overcome one main feature: no constant energy distribution throughout a year, a season, a month or even a day. Though, hydrogen could be the solution to this problem because it seems to be the ideal energy carrier and storage medium [5, 6].

Hydrogen is a promising fuel for the future mainly because it is renewable, clean and environmentally safe. Electrolysis of water is a standard commercial technology for producing hydrogen, but other technologies are in the development stage for renewable energy production, such as biomass-to-hydrogen processes, including gasification, pyrolysis and fermeation [7]. Hydrogen has high energy content per unit of mass and when it is combusted with oxygen, it is producing water as a "by-product". It releases 33.3 kWh/kg, (not included the energy contained in the residual water vapor) the same amount released from 2.5 kg of natural gas or 2.75 kg of oil [8]. However, for more practical assessments it makes more sense to refer to its energy content to a reference volume. Various methods can be used to achieve practical densities for the purpose of storage. Hydrogen can be cooled to cryogenic temperatures and stored as a liquid or, compressed to high pressures and stored as a gas at ambient temperatures [9, 10]. Nevertheless, hydrogen can also be stored at solid state. In solid state hydrogen storage, a material that can reversibly absorb or desorb hydrogen in atomic or molecular form is used to compress hydrogen (chemically or physically) to high storage densities [11]. The problem with the hydride-forming elements and compounds is that the mass of the host lattice results in a storage unit that is too heavy for practical transportation applications [12].

Many scientists all over the world are trying to come up to a solution for longer term hydrogen storage. Moreover, a lightweight high-pressure hydrogen storage vessel starts to become a reality for onboard hydrogen storage technology, while its rapid development has led the working pressure to be much higher than it used to be. As a result, the need for efficient hydrogen compressors has begun to emerge.

The two main types of hydrogen compressors are the mechanical and the non-mechanical. The first family includes mechanical piston and diaphragm compressors while the second family is mainly represented by the solid-state hydrogen compressor and the electrochemical hydrogen compressor [13].

Compression of gaseous hydrogen up to high pressure by conventional methods, represents a difficult technical problem [14]. These technical problems are related to lubrication and maintenance [15, 17] and compressor construction due to reliability of seals of moving parts [16]. In addition, mechanical compressors consume high grade electrical energy for their operations [16, 17]. Thus, it is desirable to adopt engineering solutions, which eliminate the usage of moving parts working under a hydrogen medium.

It is also well known that the hydrogen absorption-desorption plateau pressure of a metal hydride varies with the temperature according to Van't Hoff equation. Thus, the metal hydride compressors are thermally powered systems that use the properties of reversible metal hydride alloys to compress hydrogen without contamination [13]. They also provide the ability of connecting them to an outlet of an electrolyser [18] as long as the traces of dilute electrolyte, vapor and oxygen have been removed from the hydrogen produced [19].

Operation of the hydride compressor depends on heat and mass transfer in the reaction bed during the absorption and desorption processes. Though, the knowledge of these transfer characteristics is of high interest. In a theoretical study made by Muthukumar et al. [20], was presented that there are some parameters which play an important role in hydrogen compression procedure. These parameters are: heat source and sink temperatures, operating pressures, cycle time and bed parameters such as overall heat transfer coefficient, bed thickness and bed thermal conductivity

The operation of a metal hydride hydrogen compressor can be simply described as follows: First, hydrogen gets absorbed into the metal hydride at a low supply pressure and low temperature where it remains until it is exposed to a temperature increase or pressure drop when finally exits the hydride. If the temperature increase is sufficient and the final storage volume is smaller than the supply volume, the hydrogen exits the metal hydride at pressures that range from approximately 3-10 times the original supply pressure.

Multistage metal hydride hydrogen compressors, use a combination of different metal hydrides to increase the final compression ratio. Hydrogen compression is a procedure known for more than 40 years [21]. Over the last decade a great number of scientists have made great efforts in the subject of MHHC and some interesting results were found.

Muthukumar et al. [16] performed a computational study on a three - stage hydrogen compressor process and achieved a pressure ration of 28:1 while by varying the operating conditions of a single stage compressor and obtained a pressure ratio of 8.8:1 at temperature $95^0C$ of the external bath. They [22] tested MmNi4.6Al0.4 MHHC for constant and variable delivery pressures and came upon the fact that hydrogen storage pressure increases with supply pressure and heat source temperature. The increase of supply pressure has a negative effect on the efficiency of the compressor.

Talaganis et al. [23] proposed a dual – stage hydrogen compressor with a compression ratio of 3,65. Moreover, they [24] proposed a mathematical model for hydrogen compression and performed this model for two different simulation cases.

The discovery of hydrogen absorption by LaNi$_5$ and Fe-Ti alloys gave a huge thrust for studying many other different types of alloys. However, for onboard storage, these two alloys remained at the stage of prototypes, while AB$_2$ type (Laves phase), AB$_5$ type phases and Ti-based BCC alloys have become better known as hydrogen storage materials [25].

Laves phases are the largest family of binary intermetallics. The scientific interest has focused on this group of compounds because of its interesting properties as high-temperature/high-strength, superconducting, magnetic and hydrogen-storage materials [26]. They crystallize in cubic (MgCu$_2$, C15) or hexagonal (MgZn$_2$, C14 and MgNi$_2$, C36) type structures, while the only difference is the way that the same four-layered structural unit is being stacked [27].

In the current study, AB$_2$ type Zr based alloys have been synthesized by induction-levitation melting under Ar atmosphere and characterized by Rietveld analysis of the X-Ray diffraction (XRD) patterns. Furthermore, Pressure-Composition Isotherm (PCI) measurements have been conducted at 20 $^0$C, 60 $^0$C and 90 $^0$C in order to investigate the maximum pressure that can be reached from the selected alloys. Finally, SEM-EDX analysis has been conducted on polished samples of the alloys where different phase structure of the alloys was recognized. The main part of the current study is that the PCI measurements were performed in two different volumetric HIDEN's equipment with the same procedure and the aim was to compare the results found of each measurement.

## 2. EXPERIMENTAL PROCEDURE

Four different alloys were prepared under pure argon atmosphere (5N5) by using the HF-induction levitation melting method in a cold crucible and named Alloy-1, Alloy-2, Alloy-3 and Alloy-4 respectively. During the melting process, the ingots were turned over and re-melted several times (at least two) in order to ensure homogeneity. X-Ray Diffraction was performed to investigate the phase structure of experimental alloys. The diffraction data were collected at room temperature on a Philips CubiX-XRD device in the Bragg-Brentano geometry at $\lambda_{Cu-K\alpha}$ = 0.15418 nm using the following conditions of 10 < $\theta_B$ < 100 degrees and a counting step of 0.02 degrees. In order to deliver optimized information on the crystal structure modifications due to the hydrogenation procedure, Rietveld analysis (RIETICA software) has been performed on the as-cast powders as well as on the hydrogenated powders. The microstructure of each sample was investigated by using a FESEM Zeiss Ultra Plus electronic microscope. The hydrogenation dynamic and thermodynamic properties were evaluated using a volumetric high pressure Sievert's-type device. All PCI traces were recorded at 20 $^o$C and 90 $^o$C, respectively. For the activation process, all samples have been loaded into the sample holder at room temperature and under ambient atmosphere. To ensure the best degassing procedure, the samples were kept under vacuum at 350 °C for two hours. This type of alloys though, requires a high energy activation procedure in order to react effectively with hydrogen. Therefore, six-eight activation cycles have been performed using the following procedure: once the temperature of the sample has reached 350 °C following the previous out-gassing step, a pressure of 40 bars of hydrogen gas was admitted into the sample holder. Then temperature was decreased approximately up to 2 °C -5 °C and the sample was left to absorb hydrogen until equilibrium was reached. After that, the temperature was increased once more up to 350 °C and until pressure equilibrium was reached again. This procedure was repeated for 6-8 times, while the hydrogen pressure was always applied into the sample holder. Figure 1 shows the experimental procedure of such high energy activation for Alloy-2 The activation process followed for the rest samples is the same. The hydrogen gas used in the current work is 99.999% purity.

## 3. RESULTS AND DISCUSSION

### 3.1. X-Ray diffraction results

Rietveld analysis of the XRD patterns has been performed on the as-cast alloy powders as well as after pressure-composition-temperature measurements that were carried out after the activation procedure. The activation procedure is important in order to come to the highest capacity at moderate temperature. In order to achieve the highest capacity the circles of hydrogen pressure as a function of time at various temperatures have been carried out before PCTs. Figure 2, illustrates the XRD patterns of all alloys. According to the Rietveld analysis, all the alloys appear to be single phase, consisting either of the hexagonal C14 or the cubic C15 Laves phases. More specifically, Alloy-1 appear to be single phase, being consisted only of C14 laves phase, but after the PCTs, a part of the C14 phase transforms into C36. Such behavior has been reported by other authors as well [28-30]. It is observed that no significant change has taken place on the Full Width at the Half of the Maximum height if the XRD peaks after the PCTs. This means that the alloys have resistance to decrepitation after several cycles at high hydrogen pressure. It is a desirable property since metal hydride alloys are meant to work for plenty of cycles, and long life cycles are needed. All the results of the Rietveld analysis are listed in Table 1. In fact, the volume of the C15 unit cell measured here is only slightly larger than that of the ideal C36 AB2-polytype structure. Ideally it must be twice, and the very small difference as found here is explained in term of the Zr/Ti proportion varying from 1 to 3 in the A sites for C14 and C15, so the mean volume of the Friauf's polyhedron being very little larger in the cubic system than in the hexagonal system, respectively After hydrogenation-dehydrogenation processes, the grains become less than 130 nm, as it reveals by using Voight peak profile.

*Figure 1. Activation Procedure for Alloy-2*

*Figure 2. Rietveld analysis of the XRD patterns for all the synthesized alloys before and after hydrogenation measurements.*

*Table 1. Crystallographic data of the Rietveld analysis of the alloys before and after isotherms.*

*3.2. SEM-EDX analysis*

SEM-EDX analysis has been conducted on polished samples of the alloys using up to 3 μm corindon pasta. The micrographs have shown the different phase structure of the alloys, as Figure 3 shows. Figure 3a shows the micrograph for the Alloy-1. Figures 3b and 3c show the phase structure of Alloy-2 and Aloy-3 respectively and a fir-tree structure is presented in both samples. Finally, Figure 3d shows the micrograph of the Alloy-4 where a very-well defined structure is presented.

*Figure 3. SEM micrographs for all the synthesized alloys.*

It is worth to note that the Alloy-2 sample crystallize homogeneously with a typical fir-tree type structure, as most AB2 alloys having a C14 Laves phase type. Mapping of the above sample has shown the atomic distribution in the alloy, as shown in Figure 4. As it is clearly found, the Zr and Fe are at the same sites in microchemistry. The Ti and Ni distribution is at the same 'boundary-like' places while V occupies uniformly grains and boundaries in the micrometer scale.

*Figure 4. SEM chemical mapping micrographs of Alloy-2 (Zr: Red, Ti: Green, V: Blue, Fe: Turquoise, Ni: Violet).*

As it is easily observed, the $Zr_{0.5}Ti_{0.5}Fe_{1.2}Ni_{0.4}V_{0.4}$ and $Zr_{0.35}Ti_{0.65}Fe_{1.55}V_{0.45}$ alloy crystallize homogeneously with a typical fir-tree type structure, as most AB$_2$ alloys having a C14 Laves phase type, while the $Zr_{0.3}Ti_{0.7}Fe_{1.4}V_{0.6}$ alloy shows a small differentiation, probably due to the cooling process since the formula has not changed too much from that of the $Zr_{0.35}Ti_{0.65}Fe_{1.55}V_{0.45}$ sample. However, such a typical structure is absent for the $Zr_{0.75}Ti_{0.25}Fe_{1.0}Ni_{0.8}V_{0.2}$ alloy, probably related to the higher amount of both Zr and Ni. All the alloys exhibit a very good homogeneity, which is a result of the rapid solidification melting technique after quenching in the water cooled cold crucible.

*3.3. Hydrogenation Measurements*
*3.3.1. Multi-stage metal hydride compressor*
Figure 5 presents the main idea of a three–stage metal hydride hydrogen compressor. The first reactor (reactor1) is filled with the first hydride material, the second reactor

(reactor 2) is full of the second material and the third reactor (reactor 3) is filled with the third hydride. According to [16], the steps of a three – stage reactor are:

1) Absorption of low pressure hydrogen from supply tank at constant pressure (step A-B)

2) Heating of reactor 1 to external source temperature (step B-C)

3) Coupled desorption (reactor 1) – absorption (reactor 2). (Steps C-D and E-F respectively)

4) Heating of reactor 2 to external source temperature (step F-H)

5) Coupled desorption (reactor 2) – absorption (reactor 3). (Steps H-I and J-K respectively)

6) Heating of reactor 3 to external source temperature (step K-L)

7) Desorption of hydrogen from reactor 3 at high pressure at the end of compression procedure and storage into a tank (step L-M).

*Figure 5. Van't Hoff plots for operation of a three-stage metal hydride hydrogen compressor [15].*

*3.3.2 Pressure-Composition-Isotherms measurement*

PCI traces have been recorded at 20 $^{o}$C and 90 $^{o}$C, under pressure up to 100 bar. In a multi-stage metal hydride compressor, hydrogen is successively absorbed and desorbed into and out of several hydride beds. The hydrides operate between two temperature levels, 20 $^{o}$C and 90 $^{o}$C. These values are not randomly selected since 20 $^{o}$C is the temperature of the water in a city network and 90 $^{o}$C is the temperature of the water heated by waste heat at the output of an electrolyser. In the current work, the measurements were performed in two different laboratories, using a high-pressure Sievert's type apparatus. Figure 6 illustrates the isotherm measurements arriving from the first laboratory for the four-sample alloys, while Figure 7 shows the data from the measurements of the second laboratory. The results extracted from the isotherms present good reversible hydrogen storage capacities and relatively no hysteresis effect at all temperatures. The input pressure at 20 $^{o}$C should be at about 30 or 40 bars while the maximum pressure during desorption of the last alloy should be more than 150 bars. This means that the compression ratio is about five after four stages. From Figure 8, as can be seen, hydrogen pressure can be increased up to 10 times in a two stage assembly, by using such different alloy compositions.

*Figure 6.  Pressure – Composition – Temperature (PCT) Isotherms of the alloys after 8 activation cycles at 350 $^oC$ and 40 bar  (a) Alloy-1, (b) Alloy-2, (c) Alloy-3 and (d) Alloy-4. The measurements performed at laboratory No1.*

*Figure 7.  Pressure – Composition – Temperature (PCT) Isotherms of the alloys after 8 activation cycles at 350 $^oC$ and 40 bar  (a) Alloy-2, (b) Alloy-3 and (c) Alloy-3. The measurements performed at laboratory No2.*

*Figure 8.  Pressure – Composition – Temperature (PCT) Isotherms of the Alloy-2 and Alloy-4 respectively*

For the first sample, which is the lowest pressure alloy, the input conditions should be 10 bars hydrogen pressure at 20$^o$C at absorption, while the corresponding hydride can release almost 0.8 wt.% of hydrogen under  40 bar when temperature reaches 90$^o$C. This final pressure would be the input pressure at room temperature for the second sample, the highest pressure alloy. Cycling tests of the hydrogenated alloys operated under constant pressure and  temperature has showed that the kinetics are very fast as show in Figure 9 for alloy-2.

*Figure 9. Absorption and Desorption kinetics for Alloy-2.*

Maximum hydrogen uptake is reached in less than ~ 20 min (that is more than 1.6 wt.% and 1.2 wt.%, respectively) with almost 80% of the maximum uptake taking place in the first 5 min hydrogen pressure exposure. The hydrogen desorption process appears even faster at the same applied temperatures, reaching equilibrium in almost 8 min. Indeed, both these behaviors depend as well on the used system, i.e. the efficiency of the thermal transfers.

The main difference in the equilibrium pressure of plateaus, as seen experimentally here, does result of a volume effect, but more effectively on the chemical attraction to hydrogen of the metal elements. Since Zr and Ti have very similar relative electronegativities (1.4 to 1.5 in the Pauling's scale), one can anticipate that the nature and the proportion of metal elements Fe, Ni and V, plays the major role in establishing the relative stability of their respective hydrides via both the electronegativity differences and more probably the filling their 3d bands.

4. CONCLUSIONS

In the current study, AB$_2$ type Zr based alloys have been synthesized by induction-levitation melting under Ar atmosphere and characterized by Rietveld analysis of the X-Ray diffraction (XRD) patterns. Furthermore, pressure-composition isotherm (PCI) measurements have been conducted at 20 $^0$C, 60 $^0$C and 90 $^0$C in order to investigate the maximum pressure that can be reached from the selected alloys. Further microstructure investigation was performed with SEM-EDX measurements. The PCI measurements were performed in two different laboratories under the same conditions in order to validate the results. It was found that the crystal structures correspond to the C14- and C15-types of Laves phases for the Alloy-2 and Alloy-4, respectively, all samples release the whole amount of absorbed hydrogen, with almost no hysteresis effect, also the late being a very important aspect for the MHHC application. Both these favorable peculiarities (no disproportionation, almost no hysteresis) could be related to the nice homogeneity of the samples in terms of as-received microstructure. Finally, using two parent formuled-polytype of $AB_2$ compounds, the hydrogen compression ratio can reach almost 10, delivering an output pressure of more than 100 bar. The expected hydrogen transfer should be close to 1 wt.% when processing between 20 $^0$C to 90 °C using a ~180 g batch of alloy per step, thus corresponding to a little less than 20 liters $H_2$ transferred per step in about 20 min in the present but not yet optimized heat transfer conditions. When optimizing the composition of an alloy, it has to be considered whether elimination of hysteresis or maintenance of high capacity is the most important feature of the hydride. In the case of hydrides used as parts of a multistage compression system, all these parameters are very crucial in order to achieve maximum capacity. Alloys with even better PCI properties should be, and are, looked for.

**Acknowledgements**
This work was partially supported by the ATLAS-H2 IAPP European Project (Grant Agreement 251562)

Table 1. Crystallographic data of the Rietveld analysis of the alloys before and after isotherms.

| Sample Name | Space Group (No.) | Phase | Unit cell parameters (Å) | | Unit cell volume (Å³) | Phase abundance (wt%) | $R_{Bragg}$ |
|---|---|---|---|---|---|---|---|
| | | | a | c | | | |
| 1 - As milled | $P6_3/mmc$ | C14 | 4.92 | 8.02 | 505.31 | - | 5.18 |
| After PCTs | $P6_3/mmc$ | C14 | 4.92 | 8.04 | 507.53 | 49.04 | 5.12 |
| | $P6_3/mmc$ | C36 | 5 | 16.37 | 1064.99 | 50.96 | 5.93 |
| 2 - As milled | $P6_3/mmc$ | C14 | 4.93 | 8.04 | 509.15 | - | 7.82 |
| After PCTs | $P6_3/mmc$ | C14 | 4.93 | 8.04 | 509.97 | - | 5.09 |
| 3 - As milled | $P6_3/mmc$ | C14 | 4.91 | 8.01 | 503.30 | - | 5.21 |
| After PCTs | $P6_3/mmc$ | C14 | 4.92 | 8.02 | 504.86 | - | 5.40 |
| 4 - As milled | $Fd-3m$ | C15 | 6.99 | - | 341.86 | - | 5.02 |
| After PCTs | $Fd-3m$ | C15 | 6.98 | | 341.35 | | 4.98 |

**E.D. Koultoukis, E.I Gkanas, S.S. Makridis[*], C. N. Christodoulou, D. Fruchart And A.K. Stubos, "High-Temperature Activated AB$_2$Nanopowders for Metal Hydride Hydrogen Compression"**

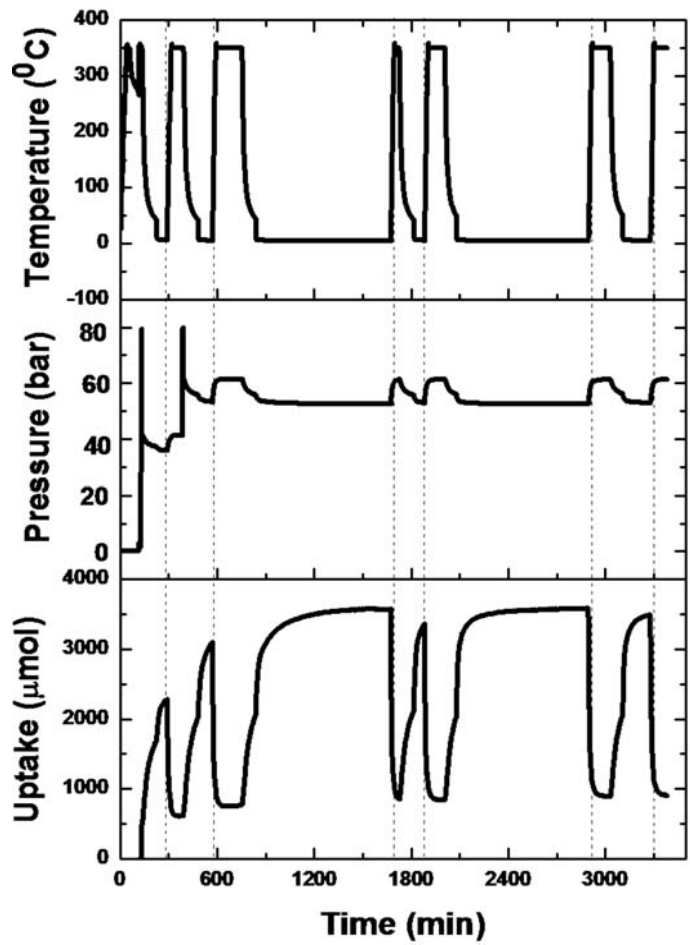

Figure 1. Activation Procedure for Alloy-2

E.D. Koultoukis, E.I Gkanas, S.S. Makridis[*], C. N. Christodoulou, D. Fruchart And A.K. Stubos, "High-Temperature Activated AB$_2$ Nanopowders for Metal Hydride Hydrogen Compression"

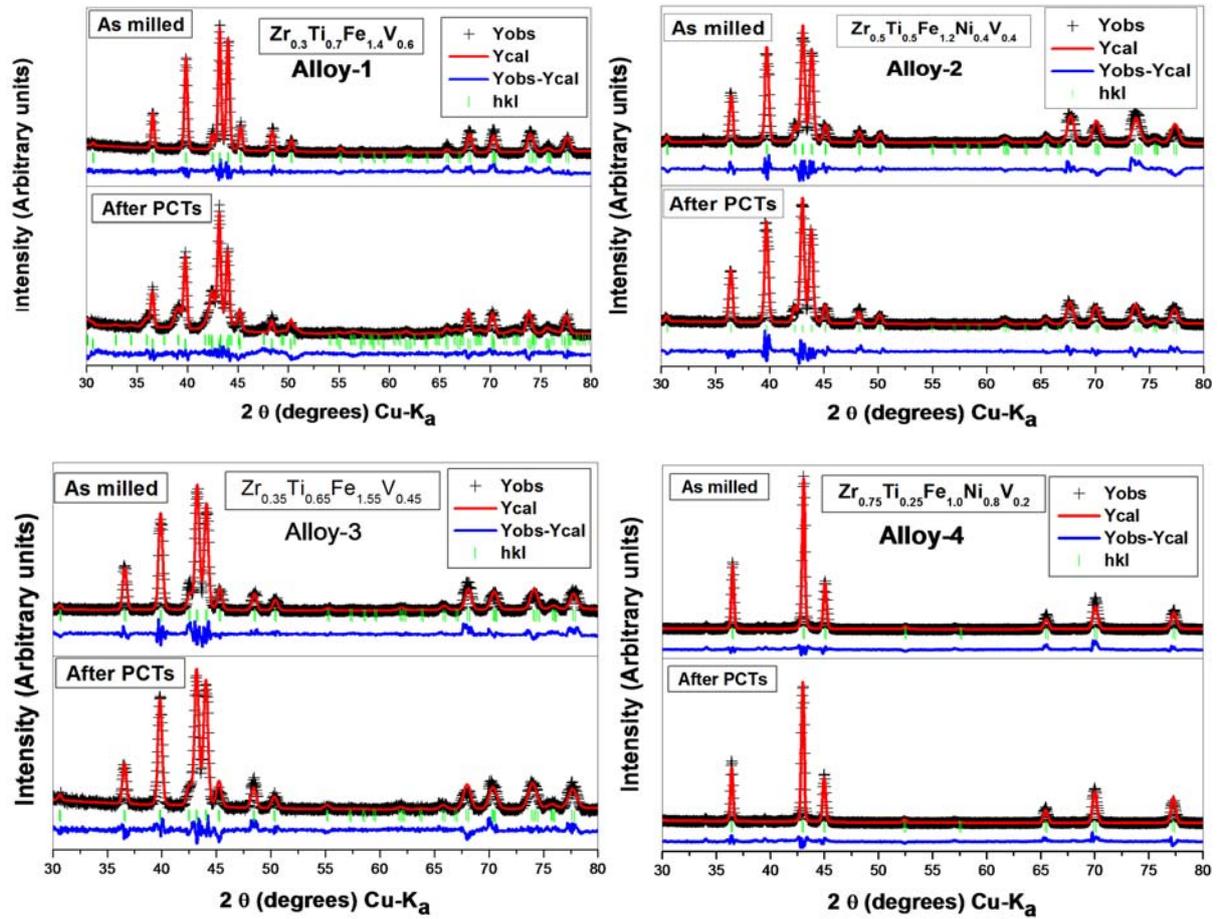

Figure 2. Rietveld analysis of the XRD patterns for all the synthesized alloys before and after hydrogenation measurements.

E.D. Koultoukis, E.I Gkanas, S.S. Makridis[*], C. N. Christodoulou, D. Fruchart And A.K. Stubos, "High-Temperature Activated AB$_2$ Nanopowders for Metal Hydride Hydrogen Compression"

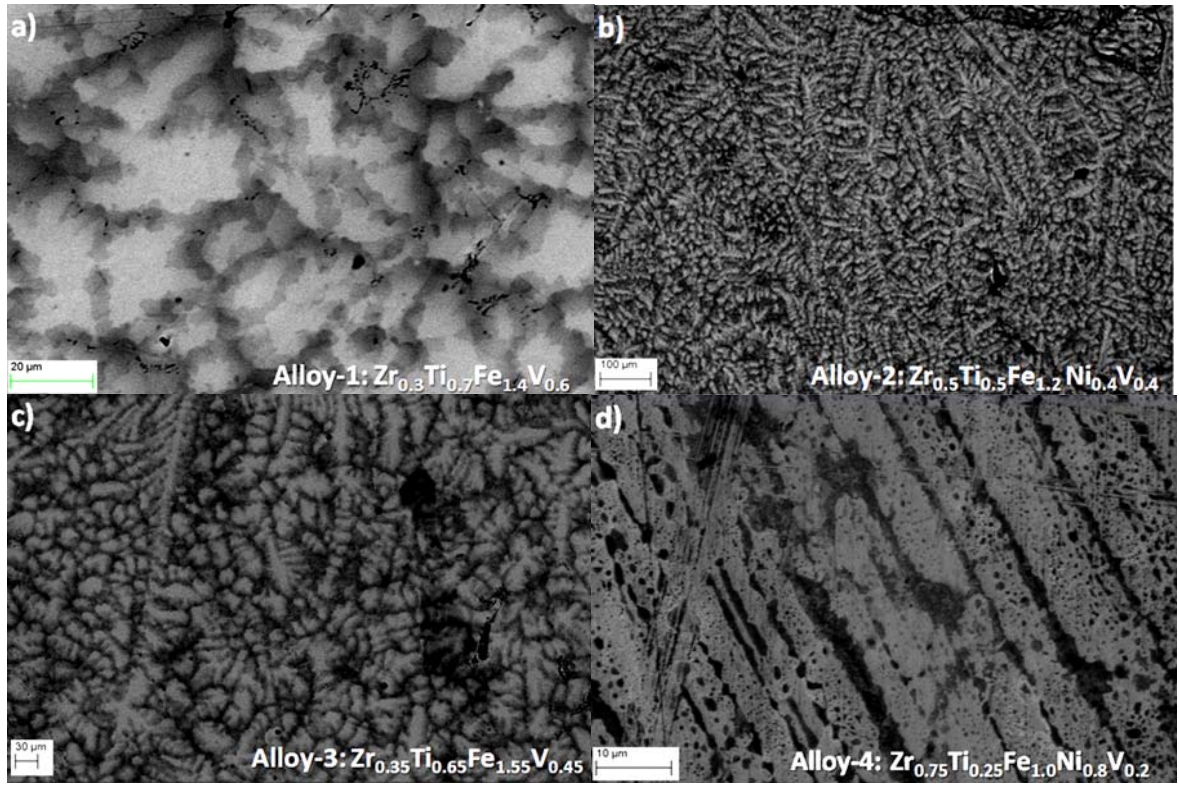

Figure 3. SEM micrographs for all the synthesized alloys.

**E.D. Koultoukis, E.I Gkanas, S.S. Makridis[*], C. N. Christodoulou, D. Fruchart And A.K. Stubos, "High-Temperature Activated AB$_2$ Nanopowders for Metal Hydride Hydrogen Compression"**

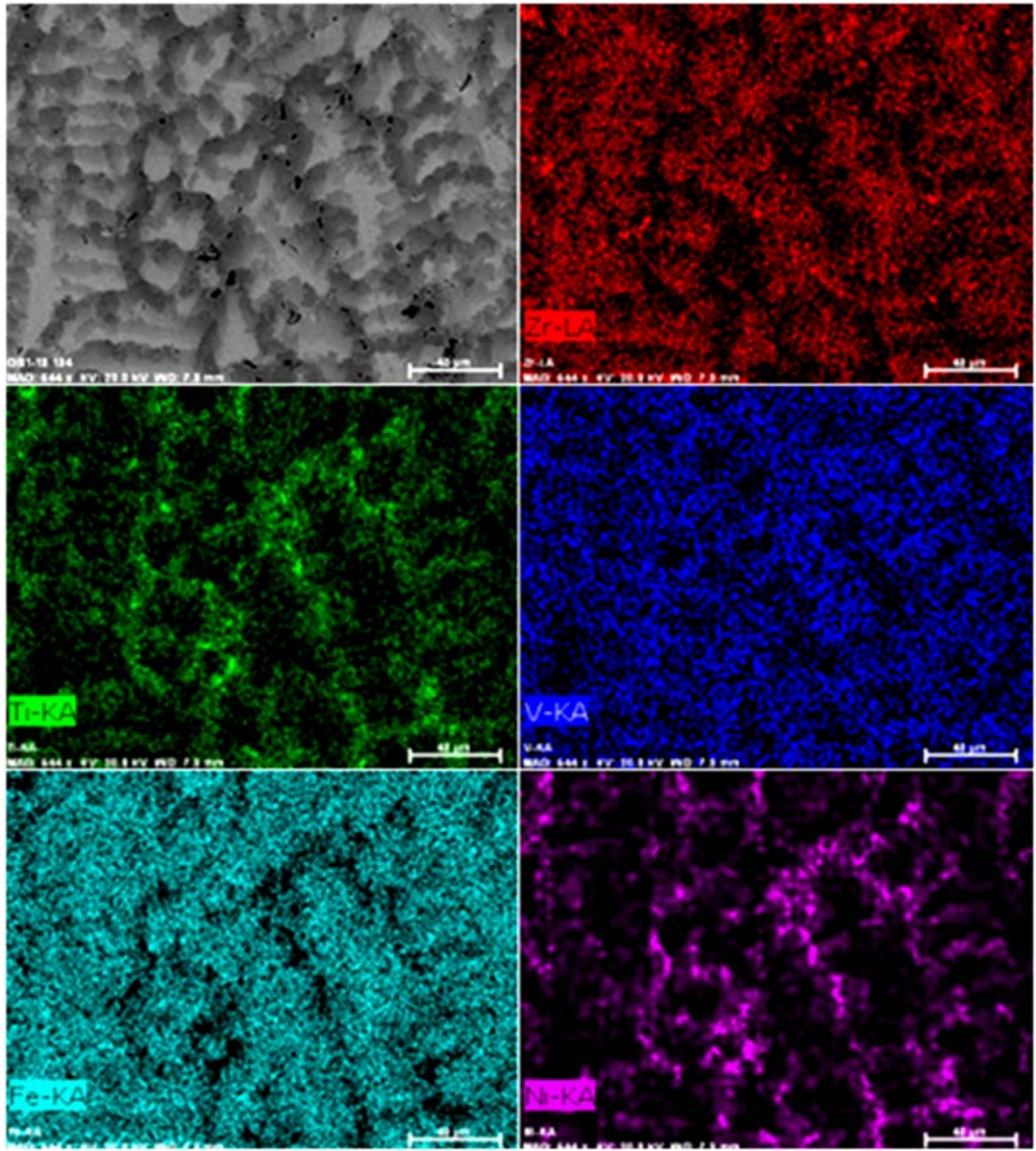

Figure 4. SEM chemical mapping micrographs of Alloy-2 (Zr: Red, Ti: Green, V: Blue, Fe: Turquoise, Ni: Violet).

E.D. Koultoukis, E.I Gkanas, S.S. Makridis[*], C. N. Christodoulou, D. Fruchart And A.K. Stubos, "High-Temperature Activated AB$_2$ Nanopowders for Metal Hydride Hydrogen Compression"

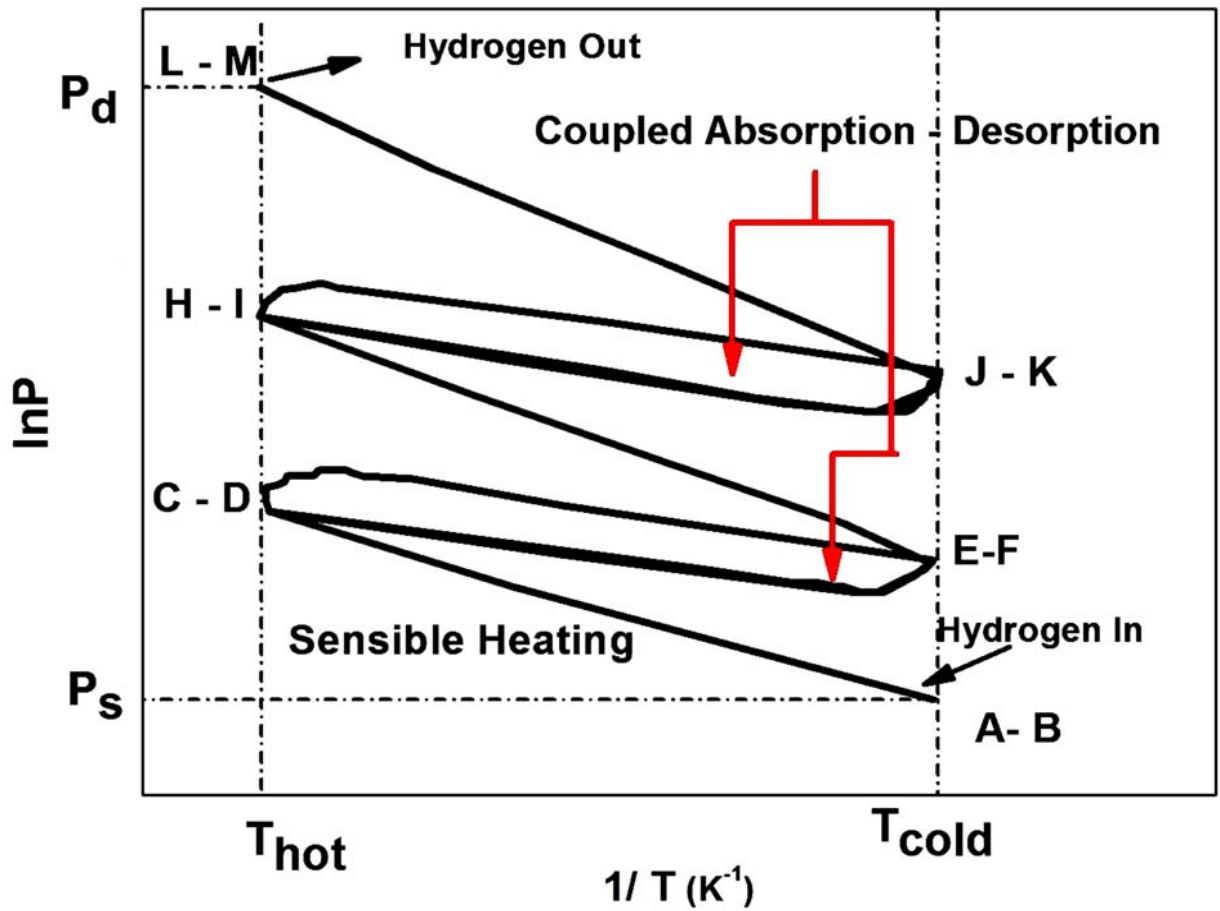

Figure 5. Van't Hoff plots for operation of a three-stage metal hydride hydrogen compressor [15].

E.D. Koultoukis, E.I Gkanas, S.S. Makridis[*], C. N. Christodoulou, D. Fruchart And A.K. Stubos, "High-Temperature Activated $AB_2$ Nanopowders for Metal Hydride Hydrogen Compression"

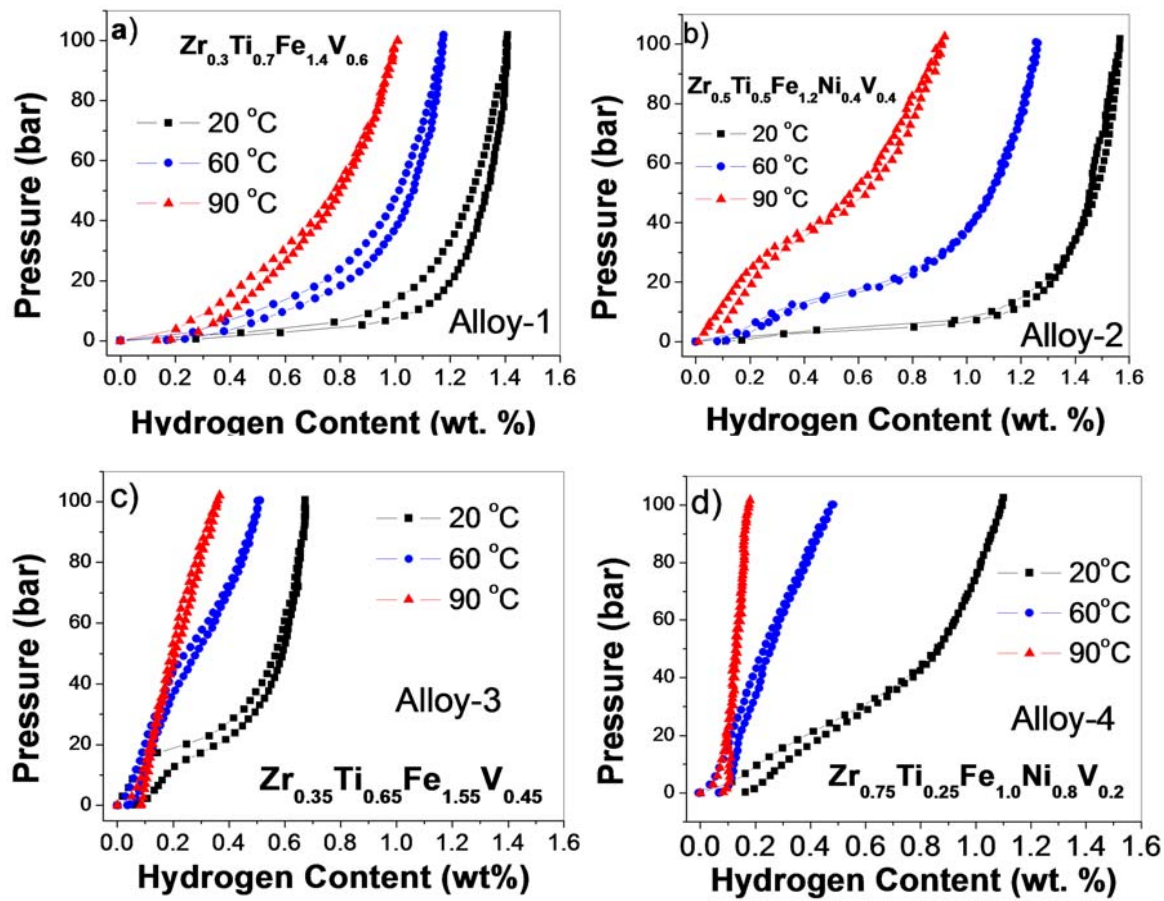

Figure 6. Pressure – Composition – Temperature (PCT) Isotherms of the alloys after 8 activation cycles at 350 °C and 40 bar (a) Alloy-1, (b) Alloy-2, (c) Alloy-3 and (d) Alloy-4. The measurements performed at laboratory No1.

E.D. Koultoukis, E.I Gkanas, S.S. Makridis[*], C. N. Christodoulou, D. Fruchart And A.K. Stubos, "High-Temperature Activated $AB_2$ Nanopowders for Metal Hydride Hydrogen Compression"

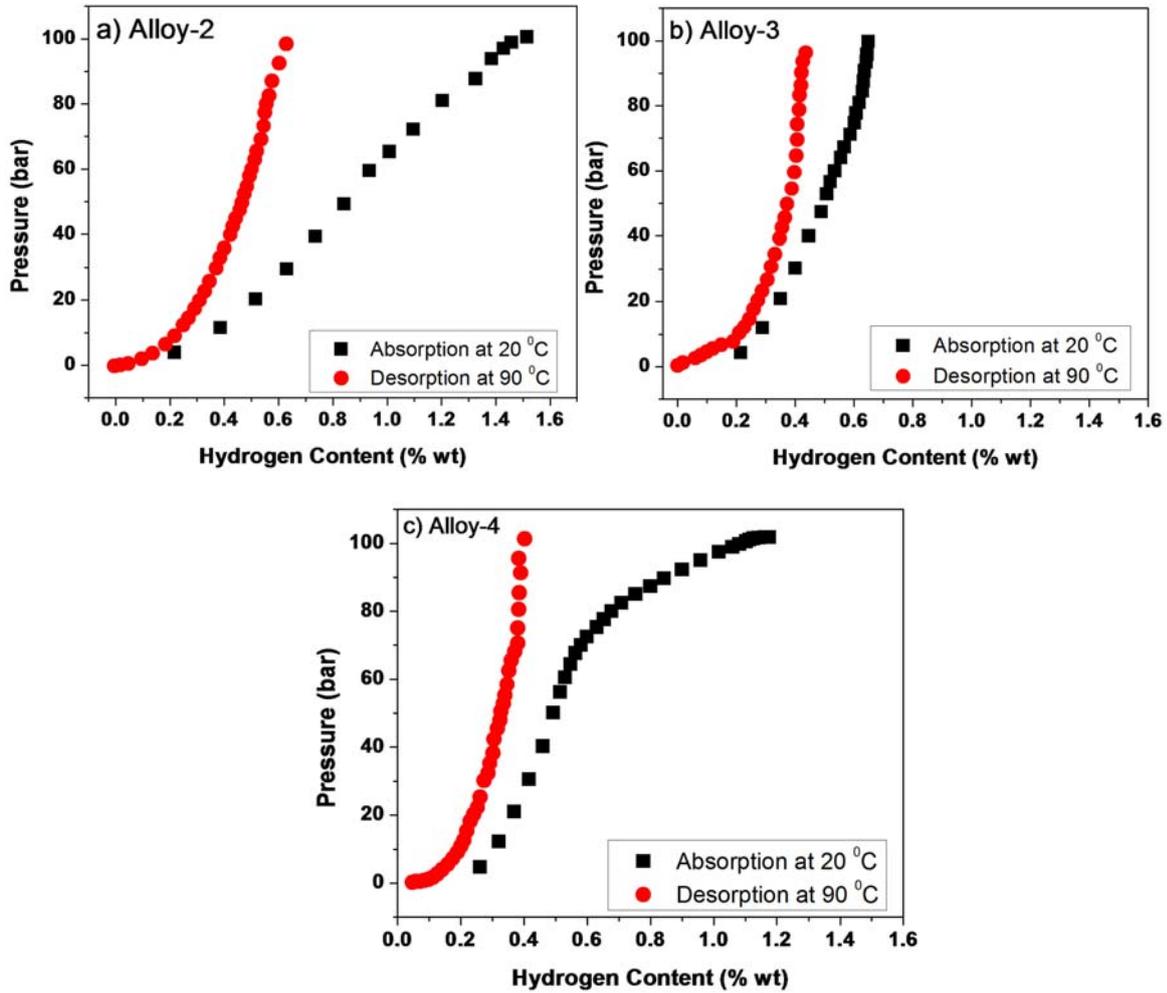

Figure 7. Pressure–Composition–Temperature (PCT) Isotherms of the alloys after 8 activation cycles at 350 °C and 40 bar (a) Alloy-2, (b) Alloy-3 and (c) Alloy-3. The measurements performed at laboratory No2.

E.D. Koultoukis, E.I Gkanas, S.S. Makridis[*], C. N. Christodoulou, D. Fruchart And A.K. Stubos, "High-Temperature Activated $AB_2$ Nanopowders for Metal Hydride Hydrogen Compression"

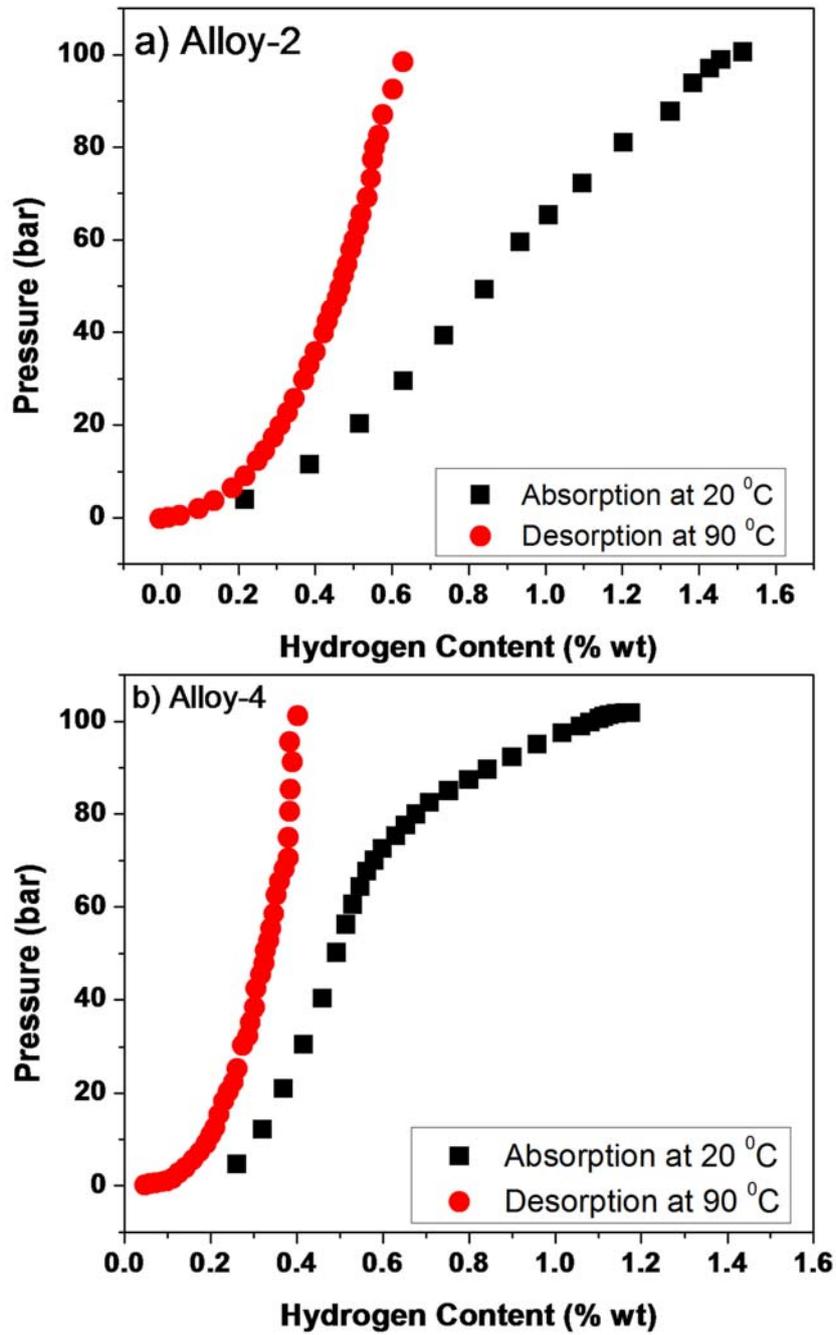

Figure 8. Pressure – Composition – Temperature (PCT) Isotherms of the Alloy-2 and Alloy-4 respectively

**E.D. Koultoukis, E.I Gkanas, S.S. Makridis[*], C. N. Christodoulou, D. Fruchart And A.K. Stubos**, "High-Temperature Activated AB$_2$ Nanopowders for Metal Hydride Hydrogen Compression"

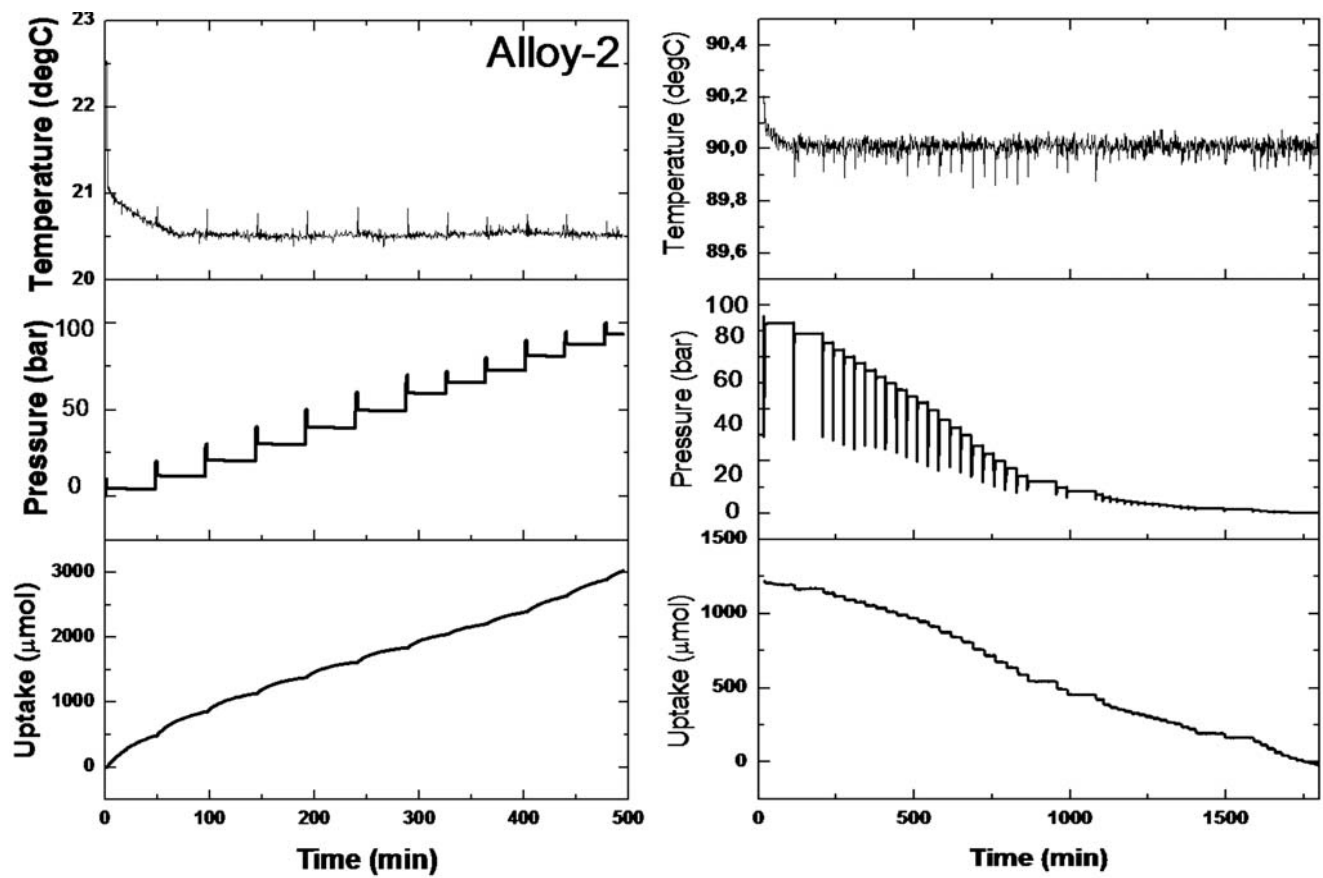

Figure 9. Absorption and Desorption kinetics for Alloy-2.

E.D. Koultoukis, E.I Gkanas, S.S. Makridis[*], C. N. Christodoulou, D. Fruchart And A.K. Stubos, "High-Temperature Activated AB$_2$ Nanopowders for Metal Hydride Hydrogen Compression"